\documentclass[a4paper,fleqn]{cas-dc}

\usepackage[numbers]{natbib}

\def\tsc#1{\csdef{#1}{\textsc{\lowercase{#1}}\xspace}}
\tsc{WGM}
\tsc{QE}
\tsc{EP}
\tsc{PMS}
\tsc{BEC}
\tsc{DE}

\begin{document}
\let\WriteBookmarks\relax
\def\floatpagepagefraction{1}
\def\textpagefraction{.001}
\shorttitle{}
\shortauthors{A. T Leblebici, S. Hassan, E. Panayirci and H. V. Poor}

\title [mode = title]{Joint Communication and Indoor Positioning Based on Visible Light in the Presence of Dimming}                      

\tnotetext[1]{This research has been supported  by the Bilateral Scientific Cooperation Program with the U.S National Science Foundation (NSF) and the Scientific and Technical Research Council of Türkiye (TUBITAK), Türkiye and, U.S NSF under Grant ECCS-2335876. It is based upon works from COST Actions CA22168 (6G-PHYSEC), supported by the European COST (Cooperation in Science and Technology) projects.}

\author[1]{A. Tarik Leblebici}
\fnmark[1]

\author[1]{Sumeyra Hassan}[%
   orcid=0000-0001-8077-2120]
\cormark[1]
\fnmark[1]
\ead{sumeyra.aldemir@khas.edu.tr}

\author[1]{Erdal Panayirci}
\fnmark[1]

\author[2]{H. Vincent Poor}
\fnmark[2]

\affiliation[1]{organization={Department of Electrical and Electronics Engineering, Kadir Has University},
                addressline={Kadir Has Caddesi}, 
                city={Istanbul},
                postcode={34083}, 
                country={Türkiye}}

\affiliation[2]{organization={Department of Electrical and Computer Engineering, Princeton University}, 
                city={Princeton},
                state={NJ},
                postcode={08544}, 
                country={USA}}             
\cortext[cor1]{Corresponding author: sumeyra.aldemir@khas.edu.tr}

\begin{abstract}
This paper proposes a joint communication and indoor positioning (JCP) system based on visible light communication (VLC) designed for high-precision indoor environments. The framework supports 2D and 3D positioning using received signal strength (RSS) from pilot transmissions, enhanced by the radical axis theorem to improve accuracy under measurement uncertainties. Communication is achieved using spatial modulation (SM) with M-ary pulse amplitude modulation (PAM), where data is conveyed through the modulation symbol and the active light-emitting diode (LED) index, improving spectral efficiency while maintaining low complexity. A pilot-aided least squares (LS) estimator is employed for joint channel and dimming coefficient estimation, enabling robust symbol detection in multipath environments characterized by both line-of-sight (LOS) and diffuse non-line-of-sight (NLOS) components, modeled using Rician fading. The proposed system incorporates a dimming control mechanism to meet lighting requirements while maintaining reliable communication and positioning performance. Simulation results demonstrate sub-centimeter localization accuracy at high signal-to-noise ratios (SNRs) and bit error rates (BERs) below $\mathbf{10^{-6}}$ for low-order PAM schemes. Additionally, comparative analysis across user locations reveals that positioning and communication performance improve significantly near the geometric center of the LED layout. These findings validate the effectiveness of the proposed system for future 6G indoor networks requiring integrated localization and communication under practical channel conditions.
\end{abstract}




\maketitle

\section{Introduction}

Joint communication and positioning (JCP) is emerging as a foundational pillar of sixth-generation (6G) wireless networks, in which communication, localization, and sensing functionalities are expected to converge into a unified framework. This convergence enables centimeter-level positioning accuracy, ultra-reliable low-latency communication (URLLC), and enhanced spectral efficiency within a shared radio infrastructure \cite{ref19}. In vehicular networks, for instance, visible light communication (VLC)-based systems have demonstrated sub-decimeter relative positioning performance in GPS-denied environments, supporting applications such as platooning, lane alignment, and collision avoidance \cite{ref16}. Intelligent Reflecting Surfaces (IRSs) have also emerged as a potential key enabler of 6G, offering controllable wireless propagation for improved beamforming and user localization in millimeter-wave (mmWave) bands \cite{ref17}. In parallel, V2X sidelink communication systems now support integrated sensing and communications (ISAC), enabling simultaneous data exchange and environmental awareness without the need for dedicated radar hardware \cite{ref18}. Recent testbeds have validated these concepts through unified waveform designs that provide encrypted communication and centimeter-level ranging using commercial off-the-shelf components \cite{ref21}.

Despite these advances in mobile and outdoor domains, indoor environments pose persistent challenges due to complex geometries and signal attenuation. Indoor positioning is increasingly critical across smart buildings, warehouses, museums, healthcare, and military facilities \cite{ref1}. Applications include asset tracking, context-aware services, and autonomous navigation. However, Global Positioning System (GPS) signals are largely ineffective indoors. The radio frequency (RF) signals used in GPS suffer severe attenuation and multipath effects when propagating through walls and enclosed structures, rendering them unsuitable for indoor localization \cite{ref22}.

To address these shortcomings, a variety of indoor positioning technologies have been explored, such as Zigbee, ultra-wideband (UWB), radio frequency identification (RFID), ultrasound, Bluetooth Low Energy (BLE), and computer vision. Each of these approaches entails distinct trade-offs. Zigbee offers low power and mesh networking but is hindered by infrastructure complexity \cite{ref2}. UWB achieves high accuracy but at the cost of synchronization and hardware complexity \cite{ref3}. RFID is low-cost but suffers from limited scalability and high error rates in dense environments \cite{ref4}. Ultrasound provides good accuracy but is sensitive to environmental conditions \cite{ref5}. BLE is power-efficient and widely adopted, yet it lacks fine-grained accuracy and is prone to interference \cite{ref6}.

Visible Light Positioning (VLP) has recently emerged as a promising candidate for high-accuracy indoor localization. VLP systems utilize ubiquitous LED lighting infrastructure for illumination and localization, with photodiodes (PDs) or image sensors acting as receivers \cite{ref23}. These systems are inherently immune to RF interference, offer high spatial resolution, and operate over the unregulated optical spectrum, aligning naturally with the goals of future JCP architectures \cite{ref20}.

VLP implementations leverage several positioning methodologies. Proximity-based systems, for example, localize users by identifying the nearest LED, suitable for low-complexity applications such as assistive navigation with 1-2 meter accuracy \cite{ref7}. Hybrid systems have also been proposed, integrating VLP with Zigbee networks to improve reliability and coverage \cite{ref8}. More advanced approaches involve triangulation using geometric properties. Angle of Arrival (AOA) techniques estimate the incidence angles of optical signals from multiple LEDs, achieving sub-decimeter precision in controlled settings \cite{ref9}. Time of Arrival (TOA) methods, which rely on delay measurements, offer centimeter-level accuracy but are sensitive to synchronization errors and multipath effects \cite{ref10}. Trilateration approaches can suffer from imperfect circle intersections due to noise and reflections; to overcome this, the radical axis theorem has been applied to enhance robustness and accuracy \cite{ref11}.

Recent innovations have pushed the boundaries of VLP performance. The perspective arcs method models circular luminaires more realistically than traditional point-source approximations, improving pose estimation under occlusion and partial visibility \cite{ref14}. In parallel, deep learning techniques such as convolutional neural networks (CNNs) and multilayer perceptrons (MLPs) have been applied to Received Signal Strength (RSS) fingerprints, enabling joint estimation of user position and orientation even in non-line-of-sight (NLOS) scenarios \cite{ref15}.

RSS-based localization remains a widely adopted technique due to its simplicity and low overhead. However, it is susceptible to environmental effects such as multipath and shadowing. Hybrid techniques combining RSS with AOA or using Received Signal Strength Difference (RSSD) have improved accuracy, achieving errors under 15 cm in practical deployments \cite{ref12, ref13}.

While most VLP studies prioritize positioning, modern wireless systems increasingly demand joint support for high-efficiency data transmission. Spatial Modulation (SM), a prominent member of the index modulation family, has attracted attention as a low-complexity, energy-efficient communication strategy. In SM, information is conveyed jointly through the index of the active transmitter, such as a selected LED and a conventional modulation symbol, thereby enhancing spectral efficiency while reducing inter-channel interference \cite{ref24}. Hence, the SM-based VLP system provides interference-free transmission with increased spectral efficiency without needing a multiplexing process at the receiving end. In addition, the positioning error of the system can be adjusted and controlled by the number of pilots employed in the estimation of the channel gains dimming coefficients without creating any data rate problem in transmission due to the higher efficiency of the SM. Consequently, in VLP systems, this approach enables simultaneous positioning and communication without increasing spectral usage or hardware complexity. By selecting LEDs based on information bits, SM naturally introduces spatial diversity, which can be exploited for user localization and data delivery \cite{ref25}. This dual-use capability aligns with the vision of multifunctional 6G networks.

Motivated by these observations, this paper proposes a unified VLP system that employs spatial modulation to achieve both high-accuracy indoor localization and robust data communication. The main contributions of this work are summarized as follows:

\begin{itemize}
\item A joint communication and positioning technique is proposed to design indoor visible light systems in the presence of dimming, addressing key challenges in 6G networks.
\item The proposed framework, which addresses key challenges in 6G networks, supports both 2D and 3D localization, with iterative refinement to improve vertical positioning accuracy yielding errors below sub-centimeter.
\item A new joint channel and dimming coefficients estimation technique is given, taking into account the Rician fading with diffuse NLOS component.
\item Spectral efficiency is enhanced via SM, where the LED index and the transmitted symbol carry data. Only one LED is activated at a time, reducing energy consumption and inter-channel interference.
\end{itemize}

\section{System Model}
\begin{figure*}[h]
    \centering  \includegraphics[width=0.8\linewidth]{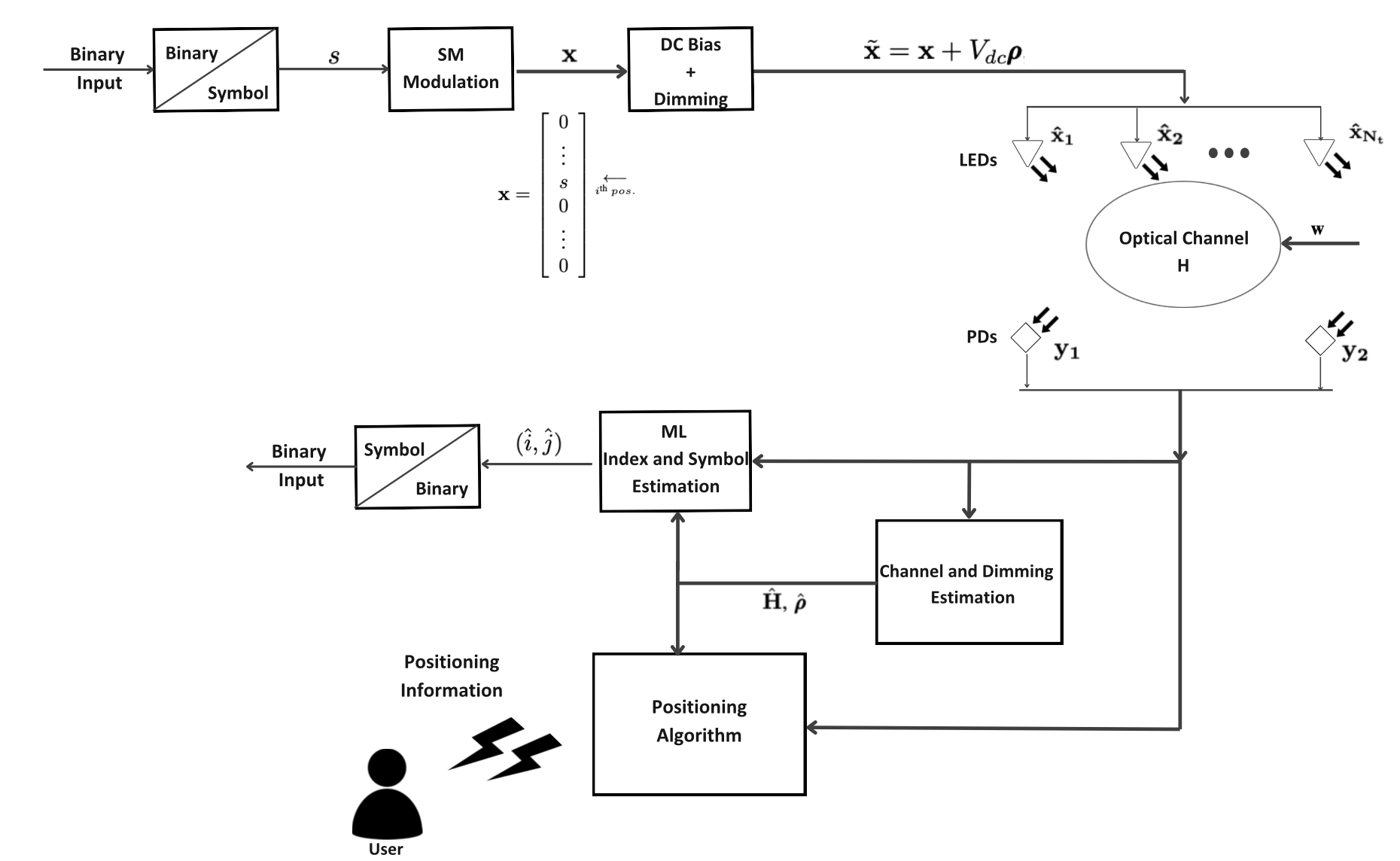}
    \caption{System model.}
    \label{fig:1}    
\end{figure*}
This paper proposes a new joint communication and positioning method based on VLC. The system employs an RSS-based radical axis approach using \(N_t\) LEDs for 3D indoor positioning. The positioning algorithm computes the RSS values for each LED at a specified resolution and compares them with real-time measurements to estimate the user’s location. The accuracy of the 3D positioning for a user equipped with \(N_r=2\) PDs improves as the resolution of the grid increases, enhancing localization precision. The optical-to-electrical (O/E) and electrical-to-optical (E/O) conversion coefficients are assumed to be unity, without any loss of generality. For communication, we incorporate an M-ary pulse amplitude modulation (PAM) SM scheme, which enhances spectral efficiency by encoding information in both the transmitted symbol and the index of the active LED. This approach significantly reduces communication complexity by activating only one LED at a time, ensuring efficient resource utilization. This joint method allows for simultaneous accurate positioning and reliable communication, with efficient VLC infrastructure. The overall system design and flow are depicted in Fig. 1.

The received signal $\mathbf{y} \in \mathbb{R}^{N_{r} \times 1}$ is expressed as
\begin{equation}
\mathbf{y}=\mathbf{H}\tilde{\mathbf{x}}+\mathbf{w},
\label{fourth_eq}
\end{equation}
where $\mathbf{H} \in \mathbb{R}^{N_{r} \times N_{t}}$ is the channel gain matrix between the LEDs and PDs, $\tilde{\mathbf{x}} \in \mathbb{R}^{N_{t} \times 1}$ is the DC-biased transmitted signal with dimming control, and $\mathbf{w} \in \mathbb{R}^{N_{r} \times 1}$ represents an additive white Gaussian noise (AWGN) vector whose each component has zero-mean and variance $\sigma_{w}^2$.

To maintain the linear operating region of the LEDs and ensure positive optical transmission, a DC bias $V_{dc}$ is introduced. In addition, a dimming control mechanism with coefficients $\rho_i$, $i = 1, 2, \dots, N_t$ is employed for each LED, allowing flexible control of brightness levels without affecting the integrity of communication. Therefore, the transmitted signal incorporating both the DC bias and dimming control is expressed as
\begin{equation}\label{xtilde1}
\tilde{\mathbf{x}} = \mathbf{x} + V_{dc} \pmb{\rho},
\end{equation}
where $\mathbf{x}$ is the modulation vector defined by
\begin{equation}\label{sdata}
\mathbf{x} = \left[ \begin{array}{ccccccccc}
0 & 0  & s & 0 & \cdots & 0 \\
\phantom{\cdots} & \phantom{1} & \underset{i^{th}pos.}{\uparrow} & \phantom{1}
\end{array} \right]^T,
\end{equation}
with $s \in \{s_1, s_2, \dots, s_M\}$  representing the M-ary PAM symbol transmitted by the  $i{\text{th}}$ LED.  Each amplitude level $s_m$ is determined by
\begin{equation}
s_m = A \cdot \left( 2m - 1 - M \right), \quad m = 1, 2, \dots, M,
\end{equation}
where A is a scaling factor. All other elements in $\mathbf{x}$ are set to zero, indicating that only the $i{\text{th}}$ LED is active for transmission during that signaling period. The dimming vector $\pmb{\rho} = [\rho_1, \rho_2, ..., \rho_{N_t}]^T$ controls the brightness of each LED, with $\rho_i \in [\rho_{\text{min}},1]$, where $\rho_{\text{min}}$ is the minimum dimming allowed.

The dimming control scheme is formulated as \(\pmb{\rho} = \pmb{\Psi}_{\text{DIM}} \pmb{\kappa}\), where \(\pmb{\Psi}_{\text{DIM}} \in \mathbb{R}^{N_t \times n_{\text{DIM}}}\) denotes the dimming matrix and \(\pmb{\kappa} \in \mathbb{R}^{n_{\text{DIM}} \times 1}\) represents the dimming coefficients for different zones or rooms. For instance, in a two-zone scenario, the dimming matrix can be structured as
\begin{equation}\label{dimmingmatrix}
\pmb{\Psi}_{\text{DIM}} = \begin{bmatrix}
1 & 1 & 0 & 0 \\
0 & 0 & 1 & 1
\end{bmatrix}^T,
\end{equation}
so that dimming levels $\kappa_1$ and $\kappa_2$ are independently applied to the respective rooms, thereby providing zone-specific illumination control while maintaining the integrity of the communication.

Considering the indoor optical transmission scheme illustrated in Fig. 1, each PD receives optical signals consisting of a line-of-sight (LOS) component and non-line-of-sight (NLOS) components due to wall reflections. The LOS channel coefficient between the $r{\text{th}}$ PD and the $t{\text{th}}$ LED is based on the Lambertian emission model, given by
\begin{equation}
	h^{ \text{\tiny LOS}}_{r,t}=\frac{(m+1)A_{\text{\tiny PD}}}{2\pi d_{r,t}^2}cos^{m+1}(\phi_{r,t})\mathbf{1}_{\Psi_c}(\phi_{r,t}),
	\label{first_eq}
\end{equation}
where \(A_{\text{\tiny PD}}\) is the PD surface area, $d_{r,t}$ is the distance between the $t$th LED and $r$th PD, and $\phi_{r,t}$ is the irradiance (and incidence) angle, assumed equal due to parallel alignment. The Lambertian emission order, $m$, is expressed as $\frac{\ln 2}{\ln(\cos(\Phi_{1/2}))}$, and $\Phi_{1/2}$ is the semi-angle at half power of the LED. The indicator function $\mathbf{1}_{\Psi_c}(\cdot)$ ensures that reception occurs only if the angle is within the PD’s field of view (FoV) and can be defined as
\begin{equation}
	\mathbf{1}_{\Psi_c}(\phi) = \Biggl\{
	\begin{array}{ll}
		1, & \text{if } |\phi| \leq \Psi_c, \\
		0, & \text{otherwise}.
	\end{array}
\end{equation}

On the other hand, the sum of a large number of diffuse reflections from multiple surfaces results in a Gaussian process with zero mean and variance $\sigma^2$, determining the NLOS component’s power. In this work, we assume {\em a priori} that the channel path gain between an LED and PD obeys a Rician distribution where $h_{r,t}$'s are Gaussian random variables with mean $\mu_{r,t}$ and variance $\sigma_{r,t}^{2}$, namely 
\begin{equation}	h_{r,t}=\mu_{r,t}+\sqrt{\sigma_{r,t}^{2}}~\breve{h},
\end{equation}
where, $\mu_{r,t}= h^{ \text{\tiny LOS}}_{r,t}$ and $\sqrt{\sigma_{r,t}^{2}}~\breve{h}$ represent the LOS and NLOS components of the Rician VLC channel in electrical domain between the $r$th PD and the $t$th LED, respectively.~$\breve{h}$ is a real-valued Gaussian random variable with zero mean and unit variance. Let
$\Omega_{r,t}=E\{|h_{r,t}|^2\}=\mu_{r,t}^2 + \sigma_{r,t}^2$
denotes the electrical power profile of the relevant Rician channel. Moreover, the Rician $ K$ factor is the power ratio of the LOS component to the diffuse component, indicating the LOS component's relative strength. In a typical $a \times b \times \ell$ indoor environment installed with VLC equipment, dividing the reflected surfaces into several small blocks, each of which satisfies the Lambertian luminance model, yields an expression of the Rician factor $K$  for  each channel between a PD  and a LED is provided by \cite{ref26} as
\begin{eqnarray}\label{Rician}
K&=&\frac{|h^{ \text{\tiny LOS}}|^2}{\sigma^2},  \nonumber\\
&=& \frac{\ell^2 \Delta s}{\rho D^4}\left[\sum_{j}^{S}\frac{\zeta\sqrt{D^2_1-(z-z_j)^2} \sqrt{D^2_2-(z_r-z_j)^2}}{D^4_1 D^4_2}\right]^{-1},
\end{eqnarray} 
where  $ \zeta= (\ell+ z_j)(z_j-\ell+5)$. $S$ is the total surface area of the reflectors, and  $\Delta s$ denotes a small reflection segment on the reflector surface located at $(x_j,y_j, z_j)$ in the indoor environment. This small area reflects the VLC signal and is then received by a PD placed at $(x_r,y_r, z_r)$. In (\ref{Rician}),  the PD's high $\ell$ and reflectivity $\rho$ are the two environmental parameters that can be assumed constant, and  $z$ denotes the third component of a given LED. $D_{1}$  is the distance between the LED and the reflected area, and  $D_{2}$ is the distance between the reflective area and the receiver. Consequently, the Rician channel gain matrix $\mathbf{H}=[h_{r,t}]$ consists of channel coefficients between the $r{\text{th}}$ PD and the $t{\text{th}}$ LED in (8) and can be expressed in terms of  $K$ as
\begin{equation}
h_{r,t}=\sqrt{\frac{K \Omega}{K+1}}+\sqrt{\frac{\Omega}{K+1}}\breve{h}.
\end{equation}

Fig. 2 shows the data frame structure utilized for positioning and communication. Each frame starts with an 8-bit start marker, followed by 400 pilot symbols dedicated to channel and dimming estimation under the Rician model. Only one LED is active during each signaling duration, following a predetermined sequence known to the receiver. This sequential activation allows the receiver to associate each received RSS measurement with the corresponding LED to estimate the user's position. Once the positioning phase is complete, the communication data bits follow. The frame concludes with a 16-bit cyclic redundancy check (CRC) for error detection and an 8-bit end marker. Since the receiver is aware of the pilot symbols, the expected RSS values, which represent the power of the transmitted symbols in the absence of noise, are also known. By comparing the measured RSS values with the expected values for each transmitter, the receiver's potential position can be estimated as the nearest point on a grid at a specific height, with each transmitter defining a unique circle of possible locations. After determining the potential locations relative to each transmitter, the radical-axis theorem can be applied to approximate the receiver's actual position.
\begin{figure}[h]
    \centering    \includegraphics[width=1\linewidth]{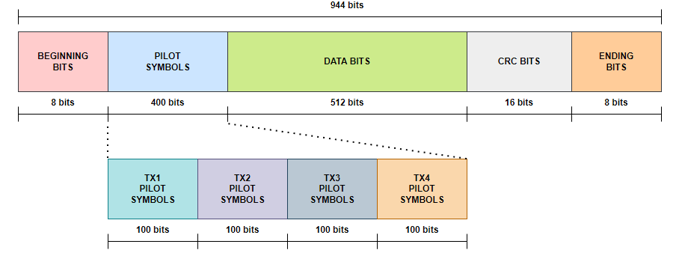}  
    \caption{Data frame structure.}
    \label{fig:2}
\end{figure}
\section{Communication Under the Presence of Dimming}
In this section, we discuss the implementation of M-ary PAM with SM within the VLC system. SM improves spectral efficiency by incorporating the random index of the active transmitter into the information-carrying process. The spectral efficiency of the system is given by
\begin{equation}
    \eta=\log_2{M} + \log_2{N_t} \,,
\end{equation}
The system's scalability is evident as increasing the number of LEDs $(N_t)$ significantly enhances the average information conveyed per symbol, offering a flexible approach to boosting spectral efficiency.

Using (1) and (2), the received signal at the PDs can be expressed as
\begin{align}\label{y_dett}
\mathbf{y} = \mathbf{H} \mathbf{x} + V_{dc} \mathbf{H} \pmb{\rho} + \mathbf{w}.
\end{align}
When the values of \(\mathbf{H}\) and \(\pmb{\rho}\) are precisely known at the receiver, the received signal (\ref{y_dett}) can be expressed as 
\begin{align} 
\tilde{\mathbf{y}}&=\mathbf{y}-V_{dc}\mathbf{H}
\pmb{\rho} \nonumber\\ &=s_i\mathbf{h}_j + \mathbf{w},
\end{align}
where $s_i$ is the transmitted M-ary PAM symbol, $\textbf{h}_j$ represents the channel gain vector corresponding to the active LED j, and $\mathbf{w}$ is the AWGN vector.

The spatial and M-PAM constellation symbols are jointly detected at the receiver using maximum likelihood (ML) detection. The ML detection aims to minimize the error between the observed received signal and all possible combinations of transmitted symbols and active transmitters as
\begin{equation}
\{ \hat{i},\hat{j} \}=\arg\min_{i,j} \|\textbf{y}-s_i\textbf{h}_j\|_F^2,
\label{eleventh_eq}
\end{equation}
where $\|\cdot\|_F$ denotes the Frobenius norm. By minimizing the squared Euclidean distance between the received signal and all hypotheses, the receiver accurately detects both the transmitted symbol and the active transmitter, ensuring robust communication even under varying dimming levels.

\subsection{Pilot-Aided Joint Estimation of Channel and Dimming Coefficients}
When the values of \(\mathbf{H}\) and \(\pmb{\rho}\) are not known at the receiver, they need to be estimated before data detection. A pilot-aided joint estimation technique is proposed, where pilot symbols embedded within transmission frames enable accurate estimation of both the channel gains and dimming coefficients. These pilot symbols provide reference signals, enabling the receiver to perform reliable channel estimation, which is critical for maintaining efficient communication in the presence of dimming.

The pilot symbol ratio  $\eta$, which represents the proportion of pilot symbols to the total number of transmitted symbols, is adjustable based on the communication environment's signal-to-noise ratio (SNR). By adjusting $\eta$, the system can dynamically optimize performance under varying channel conditions by allocating more pilot symbols when necessary, thus enhancing the accuracy of the joint estimation process.

Let \(\mathbf{x}_{p} = [0, \underset{\hspace{2mm}\Uparrow{i}}{1}, 0, \ldots, 0]^{T}\) represent the pilot vector of $N_t$ length for the $p$th pilot symbol, where $p = 1, \ldots, n_P$. Here, the index $i$ indicates the active LED during each pilot transmission period. The total number of pilot symbols employed for joint estimation is denoted by $n_P$. Since the transmitted optical signals must remain positive, an appropriate bias voltage $V_{dc}$ with a variable dimming adjustment $\rho_i, i=1,2,\ldots,N_t$ is added to each LED before transmission, leading to
\begin{equation}\label{xp}
\tilde{\mathbf{x}}_{p} = \mathbf{x}_{p} + V_{dc}\pmb{\rho},
\end{equation}
where \(\pmb{\rho} = \pmb{\Psi}_{\text{DIM}} \pmb{\kappa}\). Incorporating (\ref{xp}),  the received signal at the PDs during the pilot phase is expressed using (\ref{fourth_eq}) as
\begin{align}\label{y_received}
\mathbf{y}_{p} &= \mathbf{H}\tilde{\mathbf{x}}_{p} + \mathbf{w}_{p} \nonumber\\
&= \mathbf{H}\mathbf{x}_{p} + V_{dc}\mathbf{H}\boldsymbol{\rho} + \mathbf{w}_{p}, \quad p=1,2,\ldots, n_P \nonumber\\
&= \mathbf{X}_{p}\mathbf{h} + V_{dc}\mathbf{c} + \mathbf{w}_{p}, 
\end{align} 
where the channel vector $\mathbf{h} \stackrel{\Delta}{=} [\mathbf{h}^{T}_{1}, \mathbf{h}^{T}_{2},\ldots, \mathbf{h}^{T}_{N_r}]^{T} \in \mathcal{R}^{N_r N_t \times 1}$, $\mathbf{h}_{k} = [h_{k,1}, h_{k,2},\ldots, h_{k,N_t}]^{T}$. The matrices \(\mathbf{X}_{p}\) and \(\mathbf{c}\) are expressed as
\begin{eqnarray}
	\mathbf{X}_{p} &\stackrel{\Delta}{=}&\text{diag}\left(\mathbf{x}^{T}_{p}, \mathbf{x}^{T}_{p}, \ldots, \mathbf{x}^{T}_{p}\right) \in \mathcal{R}^{N_r \times N_t N_r},\\
	\mathbf{c}& \stackrel{\Delta}{=}& \mathbf{H}\pmb{\Psi}_{\text{DIM}}\pmb{\kappa} \in \mathcal{R}^{N_r \times 1}.
\end{eqnarray}
By expressing (\ref{y_received}) in compact matrix form, the following equation is obtained:
\begin{equation}\label{finobs}
\mathbf{y}_{p}=\mathbf{Q}_p \pmb{\xi}+\mathbf{w}_{p},
\end{equation}
where $\mathbf{Q}_{p} =\big[\mathbf{X}_{p}, V_{dc}\mathbf{H}\pmb{\Psi}_{\text{DIM}}\big]\in \mathcal{R}^{N_r \times (N_{r} N_t+n_{\text{DIM}})}$, and $\pmb{\xi}=[\mathbf{h}^{T}, \pmb{\kappa}^{T}]^{T}\in \mathcal{R}^{(N_{r} N_t+n_{\text{DIM}}) \times 1}$.

In order to jointly estimate the channel vector  $\mathbf{h}$  and the dimming vector $\pmb{\kappa}$, the received pilot signals  $\mathbf{y}_{p}$ for $p = 1, 2,\ldots,n_P$ are initially represented by the following matrix equation:
\begin{align}\label{fin_obs}
    \mathbf{Y} &= \mathbf{Q}\pmb{\xi} + \mathbf{W},
\end{align}
where $\mathbf{Y} = [\mathbf{y}^{T}_{1}, \cdots,\mathbf{y}^{T}_{n_P}]^{T}$, $\mathbf{Q}=\text{}\left(\mathbf{Q}_{1}, \cdots, \mathbf{Q}_{n_P}\right)$  and  $\mathbf{W} = [\mathbf{w}^{T}_{1}, \cdots,\mathbf{w}^{T}_{n_P}]^{T}$. The Least Squares (LS) estimation method is used to jointly estimate $\pmb{\xi}$, and the solution is provided by
\begin{align}
    \widehat{\pmb{\xi}}_{\text{LS}} &= \left(\mathbf{Q}^{T} \mathbf{Q}\right)^{-1} \mathbf{Q}^{T} \mathbf{Y},
\end{align}
where \( \widehat{\pmb{\xi}}_{\text{LS}} = \left[\widehat{\mathbf{h}}^{T}_{\text{LS}}, \widehat{\pmb{\kappa}}^{T}_{\text{LS}}\right]^{T} \).  Consequently, the LS estimates of the channel vector \( \mathbf{h} \) and the dimming vector \( \pmb{\kappa} \) are given by
\begin{eqnarray}    \widehat{\mathbf{h}}_{\text{LS}} &=& \widehat{\pmb{\xi}}_{\text{LS}}\left[1:N_t N_r\right],\\    \widehat{\pmb{\kappa}}_{\text{LS}} &=& \widehat{\pmb{\xi}}_{\text{LS}}\left[N_t N_r+1: n_{\text{DIM}}\right],
\end{eqnarray}
where the vector notation \( v[m:n] \) refers to the subvector of \( v \) containing elements from index \( m \) to \( n \). Finally, the LS estimate of \( \pmb{\rho} \) can be obtained as
\begin{align}    \widehat{\pmb{\rho}}_{\text{LS}} &= \pmb{\Psi}_{\text{DIM}} \widehat{\pmb{\kappa}}_{\text{LS}}.
\end{align} 
\section{Positioning Algorithm}

This section presents the positioning algorithm employed in the proposed system, covering both 2D and 3D positioning. In the 2D positioning scenario, where the user’s height is predetermined, the algorithm estimates the user’s 2D location by calculating the RSS values at multiple grid points on a predefined plane, as illustrated in Fig. 3. Using RSS values from multiple LEDs, the algorithm forms concentric circular patterns to identify the closest PD location by comparing measured and pre-calculated RSS distributions. Since the user’s height is known, the process is simplified to estimate only the horizontal coordinates, making the positioning more efficient. When the height is unknown, the 3D positioning algorithm iterates over different height values to estimate the 2D position at each level. The system performs 2D positioning for each PD at varying heights, creating a set of potential 3D locations. The height is then determined by matching the distance between the two estimated PD positions to their known physical separation. By iteratively estimating the user’s position at different height levels, the system accurately determines the full 3D location while properly considering how signal strength varies with height.
\begin{figure}[t]
    \centering   \includegraphics[width=1\linewidth]{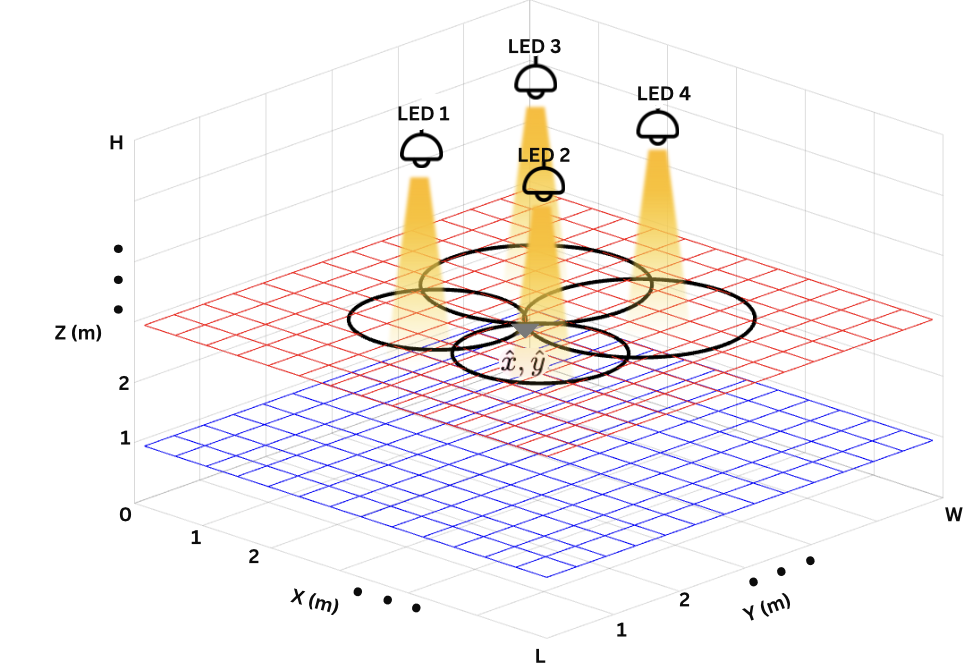} 
    \caption{LED setup and receiver plane used for 2D and 3D positioning.}
\end{figure}
\subsection{2D Positioning}
Before initiating 2D positioning, the receiver plane is divided into grid points, $g$, and the LEDs are positioned at a known height. The room dimensions are considered to be $L \times W \times H$. Each point on this grid represents a potential PD location, allowing the system to evaluate candidate positions across the plane.
It is important to note that in practical scenarios, RSS-based positioning systems are susceptible to noise and multipath propagation effects. Therefore, assuming the transmitted signal $s$  has average unit power, the total electrical power received at each grid point consists of three components: LOS signal, NLOS signals, and ambient noise powers denoted as
\begin{equation}   
	P_{\text{total}}= P_{\text{LOS}}+ P_{\text{NLOS}}+ P_{\text{AMBIENT}},
\end{equation}
where from (1) and (8) it follows that $ P_{\text{LOS}}=h_{r,t}^{\text{LOS}}$, $P_{\text{NLOS}}=\sigma_{r,t}^2$ and $ P_{\text{AMBIENT}}=\sigma^2_{w}.$  Consequently, for each LED, the RSS values at the actual PD location $g$, with $g=1,2, \cdots, G,$ is computed as
\begin{equation}\label{RSS_equation}
RSS_g = \frac{1}{n_P} \sum_{p=1}^{n_P} (s_p \hat{h}_{r,t}^{(g)}+ w_{r,t} )^2,
\end{equation}
for \(t = 1, 2, \dots, N_t\) and \(r = 1, 2\),  where \( n_P \) denotes the number of transmitted pilot symbols, \( s_p \) is the transmitted pilot symbol, and \( \hat{h}_{r,t}^{(g)} \) represents the estimated Rician-distributed channel coefficient between the \(t\)th active LED and the \(r\)th PD at the \(g\)th grid point, as used in (\ref{RSS_equation}). In addition, \( w_{r,t} \) denotes the additive white Gaussian noise (AWGN) with zero mean and variance \( \sigma_w^2 \). The positions of each grid point and the corresponding RSS value are stored for use in subsequent algorithmic processes. By leveraging the estimated channel coefficients, distinct RSS distributions are generated for each LED, resulting in characteristic circular patterns, as illustrated in Fig. 4. These patterns depict RSS variations over the receiver plane at a height of z = 0, with the LEDs located at a height of 2.15 meters. Due to the Lambertian light propagation model, the RSS values form concentric circles, where points with equal RSS values align.
\begin{figure}[t]
    \centering    \includegraphics[width=1\linewidth]{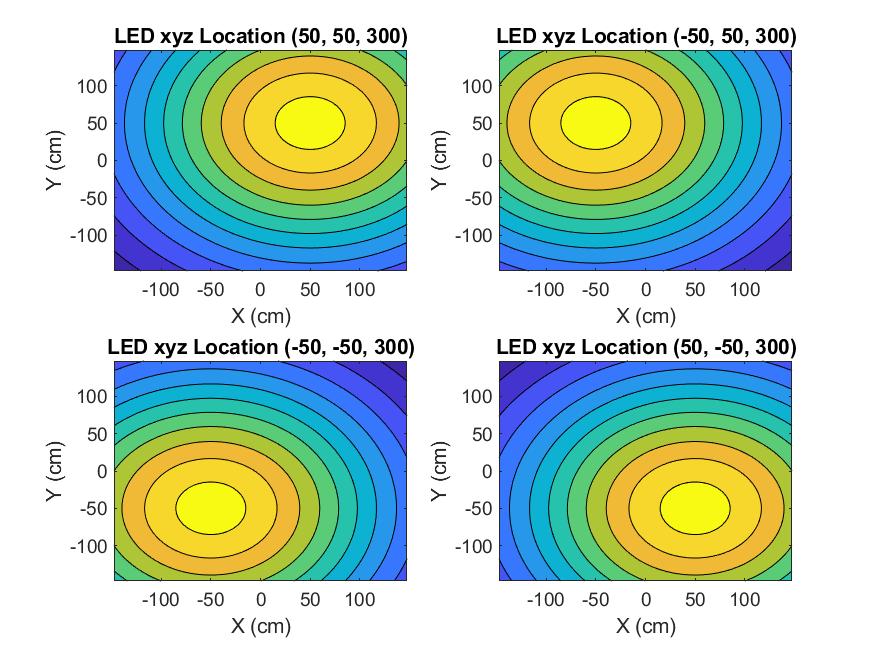}   
    \caption{RSS distributions on the receiver layer (z = 0) for different transmitter locations.}
    \label{fig:3}
\end{figure}
\begin{figure}[b]
    \centering   \includegraphics[width=0.9\linewidth]{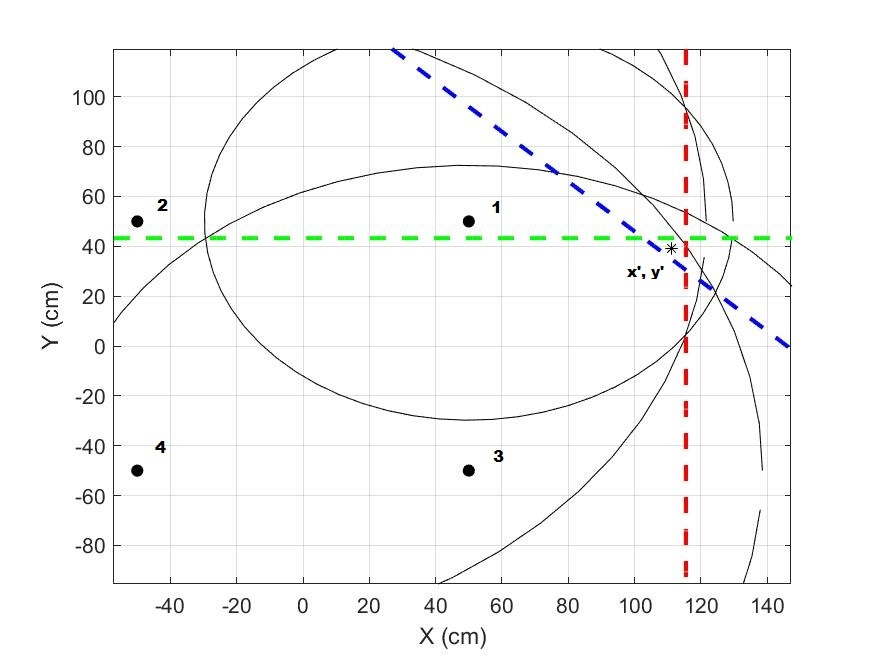}      \caption{Illustration of the radical axis theorem.}
\end{figure}

The calculated RSS values are compared with pre-computed distributions, yielding circles representing regions of equal RSS, each centered on the corresponding LED, as shown in Fig. 5. The radical axis theorem is applied to estimate the PD’s 2D location. The radical axis defines the locus of points where the power distances between two circles are equal, even if the circles do not intersect. Dashed lines in Fig. 5 represent the radical axes calculated between pairs of LEDs, each line corresponding to a set of potential PD locations based on relative RSS measurements. The intersection of these radical axes provides the estimated position of the two PDs at coordinates $(\hat{x}_k, \hat{y}_k), k=1,2$, offering a precise and robust location estimate. Additionally, even if the radical axes do not intersect at a single point, a point resulting in a good enough estimation can still be achieved by taking the centroid of the triangle lines created. By leveraging the radical axis method, the system achieves improved positioning accuracy even under multipath distortion or ambient noise conditions.

\subsection{3D Positioning}
The 3D positioning algorithm requires two PDs separated by a known distance $d$. The height values, indexed from $1$ to $n$, are stored in a row vector \(\mathbf{z} = [z_1, z_2, \dots, z_n]^T\), representing the different discrete height levels considered within the room. The 2D positioning results for each PD at different heights are stored in matrices, denoted as $\mathbf{P_{\text{PD}_1}}$ and $\mathbf{P_{\text{PD}_2}}$, representing the estimated 2D coordinates at each height:
\begin{equation}
\mathbf{P_{\text{PD}_k}} = \begin{bmatrix}
    x_{k,1} & y_{k,1} \\
    x_{k,2} & y_{k,2} \\
    \vdots & \vdots \\
    x_{k,n} & y_{k,n}
\end{bmatrix}, \quad k = 1, 2.
\end{equation}
The difference between the estimated 2D positions for the two PDs is calculated as follows:
\begin{equation}
\Delta \mathbf{P} = \mathbf{P_{\text{PD}_1}} - \mathbf{P_{\text{PD}_2}} = \begin{bmatrix}
    x_{1,1} - x_{2,1} & y_{1,1} - y_{2,1} \\
    x_{1,2} - x_{2,2} & y_{1,2} - y_{2,2} \\
    \vdots & \vdots \\
    x_{1,n} - x_{2,n} & y_{1,n} - y_{2,n}
\end{bmatrix}.
\end{equation}
Here, $\Delta \mathbf{P}$ matrix contains the differences in the $x$- and $y$-coordinates of the two PDs at each height level. Next, the squared Euclidean distances between the two PDs at each height are computed as
\begin{equation}
\sqrt{\text{diag}(\Delta \mathbf{P} \Delta \mathbf{P}^T)} = \begin{bmatrix}
    \sqrt{(x_{1,1}-x_{2,1})^2 + (y_{1,1}-y_{2,1})^2} \\
    \sqrt{(x_{1,2}-x_{2,2})^2 + (y_{1,2}-y_{2,2})^2} \\
    \vdots \\
    \sqrt{(x_{1,n}-x_{2,n})^2 + (y_{1,n}-y_{2,n})^2}
\end{bmatrix}.
\end{equation}
This gives us the estimated distance between the two PDs at each height. The correct height corresponds to the row index where the estimated distance most closely matches the known physical distance $d$. This is achieved by minimizing the norm of the difference between the estimated distances and $d$ as given:
\begin{equation}
\hat{n} = \arg \min_n \| d \cdot \mathbf{1}_n - \sqrt{\text{diag}(\Delta \mathbf{P} \Delta \mathbf{P}^T)} \|.
\end{equation}
Once $\hat{n}$ is determined, the corresponding height value is selected from $\mathbf{z}$, and the associated 2D coordinates from $\mathbf{P_{\text{PD}_1}}$ and $\mathbf{P_{\text{PD}_2}}$ are used to finalize the 3D position estimation of the user.

\section{Computer Simulations}
\begin{figure*}
    \centering    \includegraphics[width=1\linewidth]{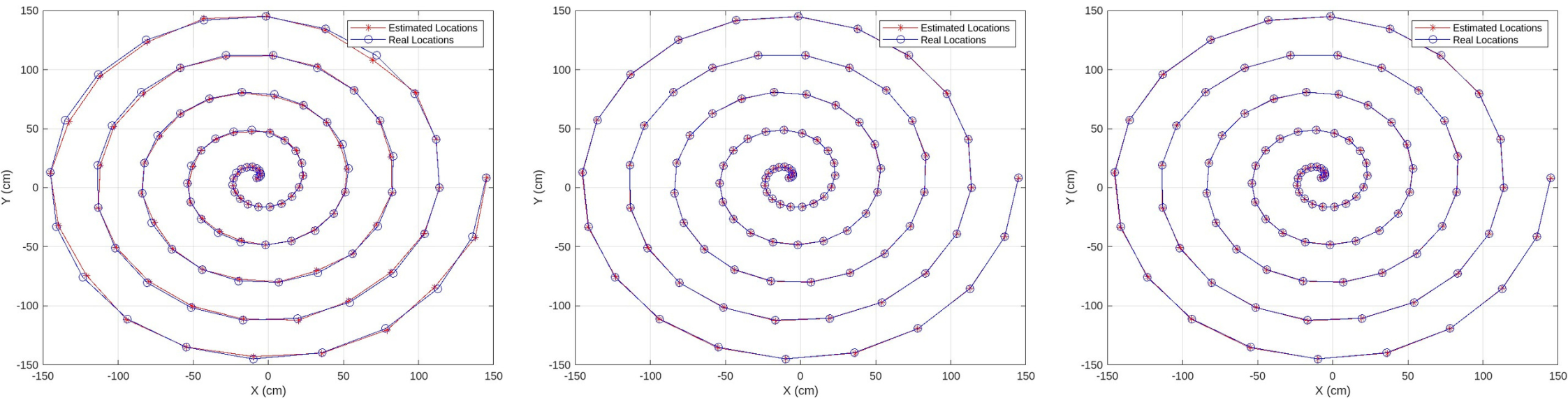} 
    \caption{2D Positioning performance at 20 dB, 50 dB and 80 dB SNR.}
    \label{fig:7}
\end{figure*}
\begin{figure*}
    \centering    \includegraphics[width=0.8 \linewidth]{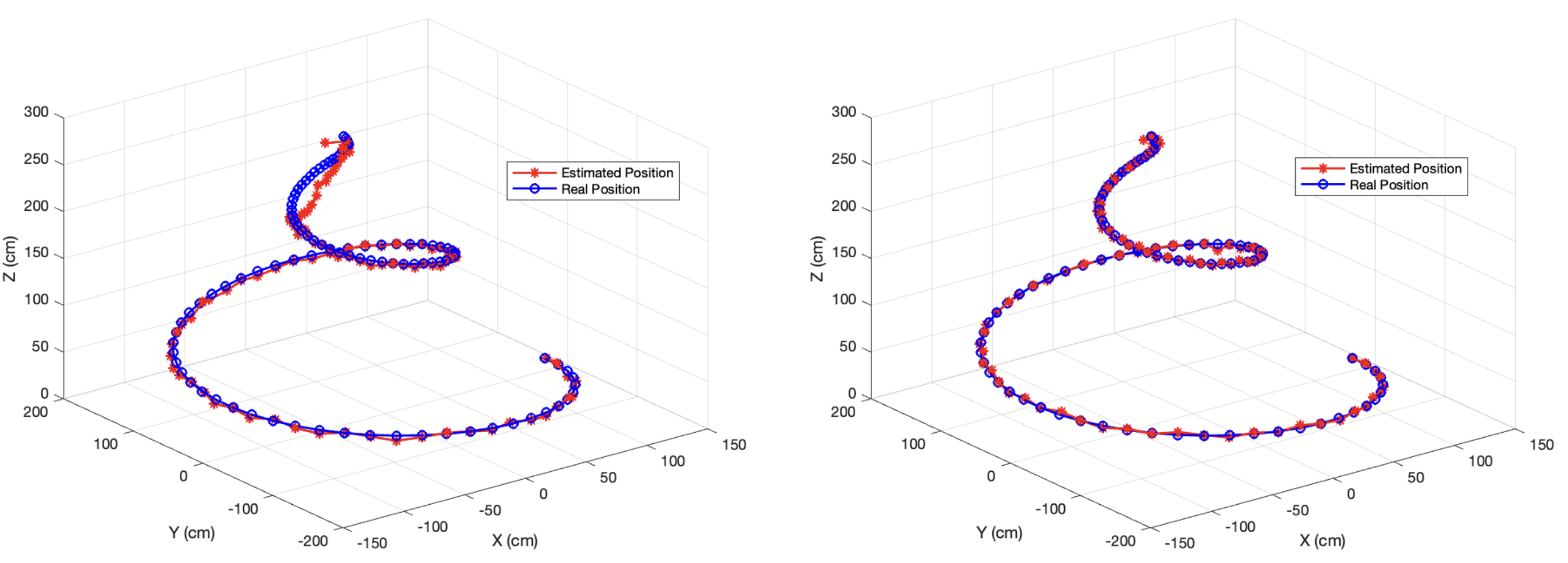} 
    \caption{3D Positioning performance at 20 dB and 80 dB SNR.}
    \label{fig:8}
\end{figure*}
This section presents the computer simulation results for the proposed VLC-based positioning method, conducted in a room with dimensions of 300cm \( \times \) 300cm \( \times \) 300cm, with \( N_t = 4 \) LEDs and \( N_r = 2 \) PDs. Each LED has a transmit power of \( P_t = 20 \text{W} \), a half-power semi-angle of \( \Phi_{1/2} = 60^\circ \), and a Lambertian emission order of \( m = 0.647 \). The PDs, located at varying heights, are characterized by an optical filter transmission factor of \(T_s = 1\), a refractive index of \(n = 1.5\), a field of view (FoV) half-angle of \(\Psi_c = 60^\circ\), and an optical concentrator gain of \(g = 2.6\). The simulations evaluate the positioning and communication performance across varying SNR levels, particularly the influence of the receiver's distance from the room center on the system performance. The LED locations are defined as
\begin{equation}
\mathbf{LED}_{(x,y,z)} = \begin{bmatrix}
50 & 50 & -50 & -50 \\
50 & -50 & 50 & -50 \\
300 & 300 & 300 & 300
\end{bmatrix},
\label{eq:led_positions}
\end{equation}
where each column represents the coordinates \((x, y, z)\) of a respective LED. The simulations use these locations to assess positioning and communication performance across varying SNR levels, comprehensively evaluating the system’s effectiveness in 2D and 3D environments. Two distinct scenarios are considered to evaluate the proposed method's 2D and 3D positioning performances. The first scenario involves a 2D spiral path on a plane, while the second uses a 3D spiral path that gradually decreases in radius as the receiver moves upward. The spatial grid resolution is set to 5 cm, directly impacting the positioning accuracy; a finer grid provides more precise localization at the cost of increased computational complexity. The simulations comprehensively assess the system's joint communication and positioning capabilities under various SNR and environmental conditions, including NLOS components modeled using a Rician distribution.

In the 2D positioning scenario, the impact of SNR on localization accuracy is illustrated in Fig.~\ref{fig:7}, where the real and estimated positions are plotted for SNR levels of 20 dB, 50 dB, and 80 dB. As expected, the accuracy of the estimated trajectory improves significantly with increasing SNR. At 20 dB, the estimated positions deviate noticeably from the real position, reflecting a substantial localization error. In contrast, at 50 dB and above, the estimated trajectory aligns closely with the actual one, demonstrating the algorithm's effectiveness under higher SNR conditions. The results highlight the algorithm's ability to deliver precise 2D localization performance when supported by sufficiently strong signal quality.

In the 3D positioning scenario, shown in Fig. \ref{fig:8}, the real and estimated positions are presented for 20 dB and 80 dB SNR. Similar to the 2D scenario, the accuracy improves as the SNR increases. At 20 dB, the estimated trajectory deviates more, particularly at higher vertical positions of the spiral. However, at 80 dB SNR, the estimated positions almost perfectly trace the real spiral path with minimal error, demonstrating the effectiveness of the proposed method in 3D environments.
\begin{figure}[b]
    \centering  \includegraphics[width=1\linewidth]{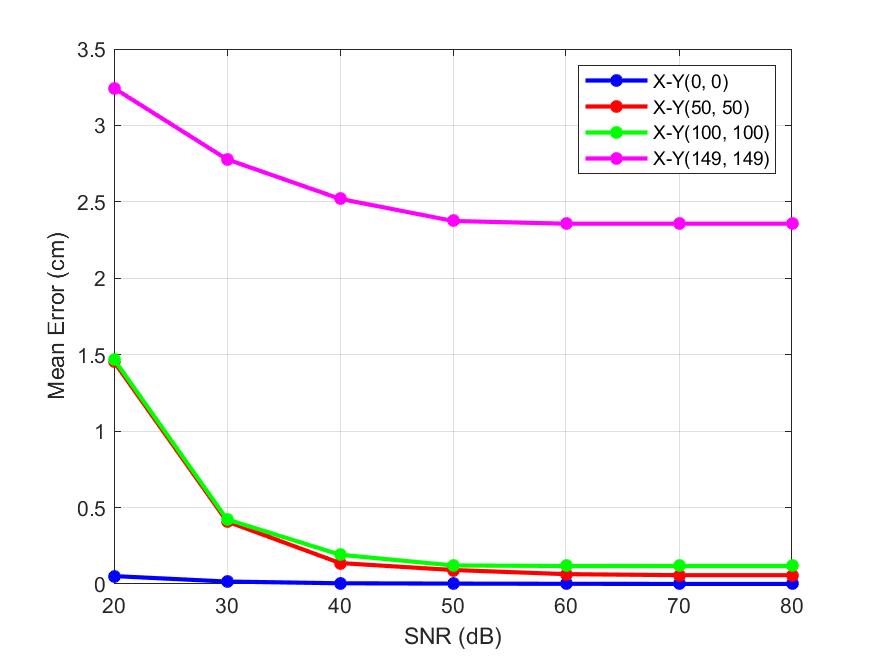}
    \caption{Average performance of 2D positioning.}
    \label{fig:9}
\end{figure}
\begin{figure}[b]
    \centering    \includegraphics[width=1\linewidth]{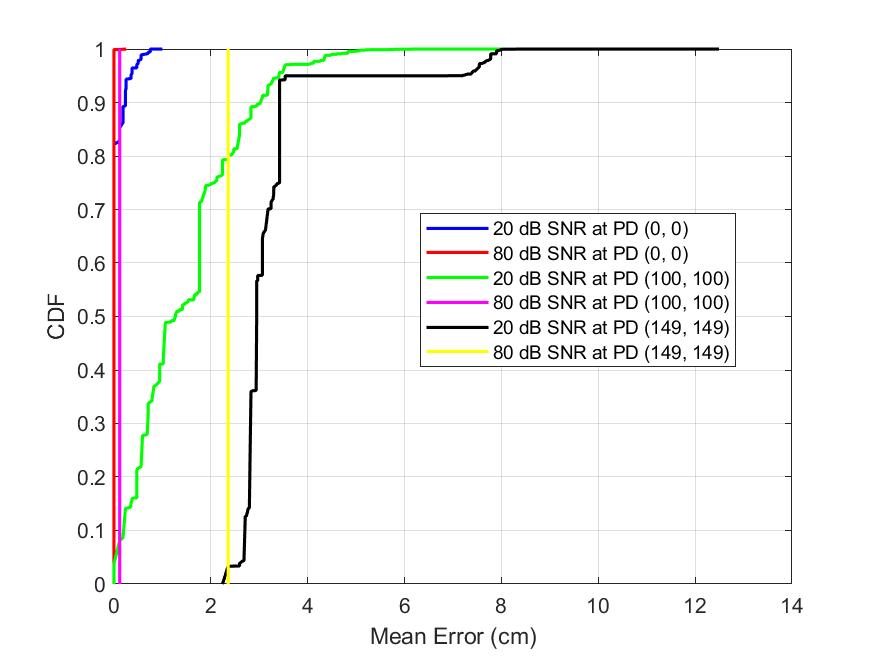}
    \caption{Emprical CDF for 2D positioning at different locations.}
    \label{fig:14}
\end{figure}

In Fig.~\ref{fig:9}, the average 2D positioning performance is illustrated as a function of SNR for four receiver locations: the room center (0, 0) and three progressively farther positions toward the corner (50, 50), (100, 100), and (149, 149). The results show that the mean positioning error decreases significantly as SNR increases, with the most notable improvement observed between 20 dB and 40 dB. At locations near the center of the room, the system achieves sub-centimeter accuracy across the entire SNR range, with mean errors remaining below 0.1 cm even at 20 dB. In contrast, for corner locations such as (149, 149), the positioning error starts above 3 cm at 20 dB and gradually decreases to approximately 2.3 cm at higher SNR values. This performance gap is attributed to the increased impact of diffuse NLOS components at the edges and corners of the room, where the LOS contribution is weaker. These observations confirm that the proposed 2D positioning method is robust under moderate-to-high SNR conditions and performs most accurately when the receiver is closer to the center of the LED layout.

The empirical cumulative distribution functions (CDFs) curves in Fig.~\ref{fig:14} show the effect of SNR and receiver location on 2D positioning accuracy. At the center of the room \((0, 0)\), the system achieves near-perfect accuracy even at 20 dB, with errors consistently below 0.2 cm. As the receiver moves toward the corner \((100, 100)\), the error spread increases at 20 dB but improves significantly at 80 dB. In contrast, at the far corner \((149, 149)\), positioning errors exceed 10 cm at 20 dB and remain around 2-3 cm even at 80 dB. These results highlight that while higher SNR enhances performance, receiver location strongly influences accuracy due to the dominance of NLOS components near room edges.
\begin{figure}[t]
    \centering   \includegraphics[width=1\linewidth]{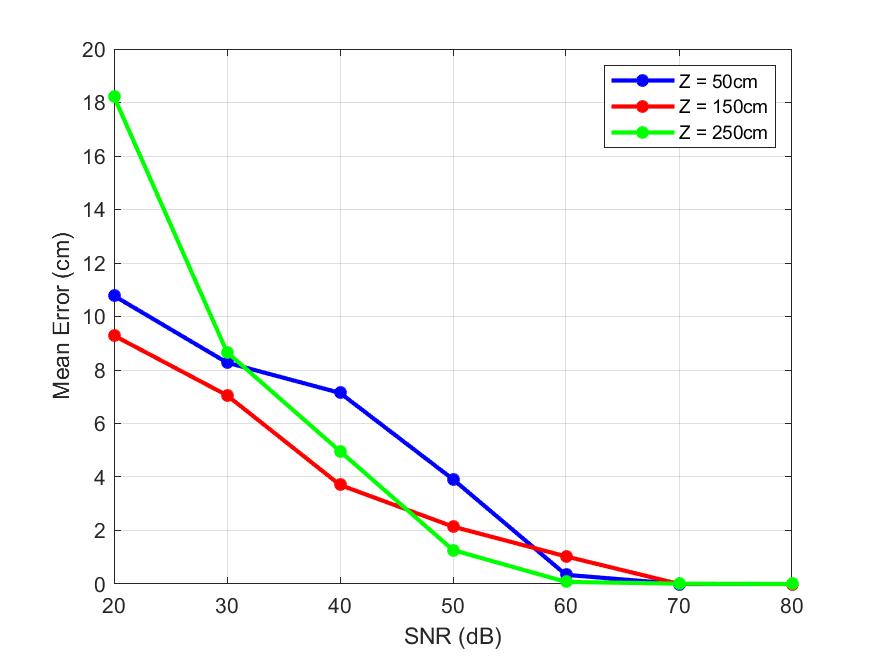}
    \caption{Average performance of 3D positioning.}
    \label{fig:10}
\end{figure}

In Fig.~\ref{fig:10}, the average performance of the proposed 3D positioning method is evaluated in terms of SNR versus mean positioning error under varying height levels. The simulation is conducted for a fixed horizontal location $(X,Y) = (100,100)$ cm while varying the vertical position $Z \in \{50, 150, 250\}$ cm. Each curve represents the average error computed over $10^5$ Monte Carlo runs to ensure statistical robustness. Similar to the 2D results, the positioning accuracy improves significantly as the SNR increases. At low SNR levels (e.g., 20 dB), the mean error ranges between 9 cm and 18 cm, depending on the height. Notably, the error increases with vertical distance, indicating that higher positioning planes experience more severe accuracy degradation due to geometric dilution and increased path loss propagation. As the SNR exceeds 50 dB, the mean error converges to sub-centimeter levels across all heights, reflecting the system’s ability to compensate for height-induced challenges under high-SNR conditions. The results confirm that the proposed approach achieves reliable and precise 3D localization, even under realistic propagation conditions, including diffuse NLOS components.
\begin{figure}[t]
    \centering    \includegraphics[width=1\linewidth]{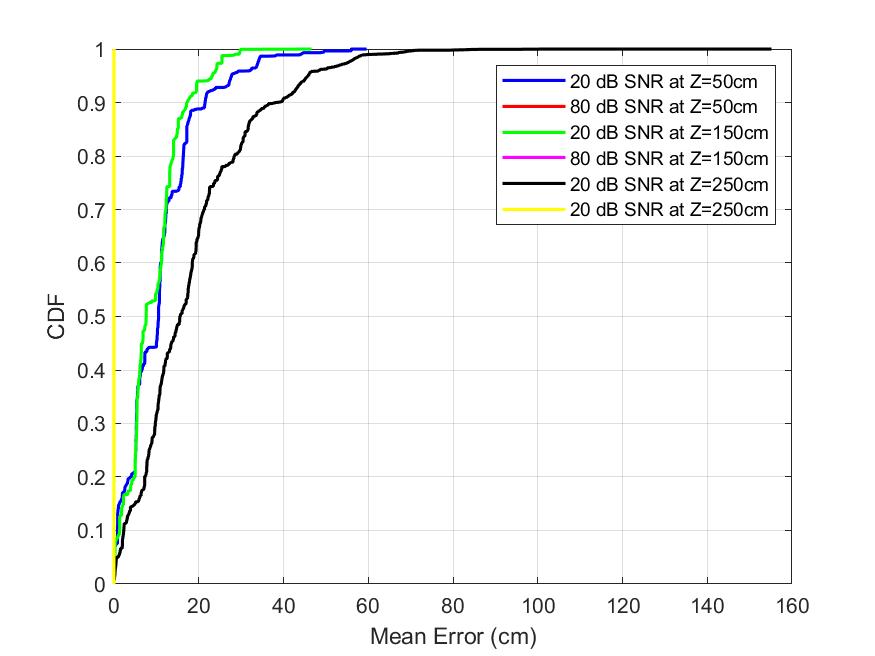}
    \caption{Emprical CDF for 2D positioning at different heights.}
    \label{fig:15}
\end{figure}

Similarly, Fig.~\ref{fig:15} presents the empirical CDFs of the 3D positioning error for various height levels and two distinct SNR values: 20 dB and 80 dB. The curves highlight the impact of both SNR and vertical position on localization performance. At 80 dB, the system achieves highly accurate localization, with errors falling below 1 cm across all heights. In contrast, at 20 dB SNR, the positioning error increases noticeably, especially at higher vertical positions. For instance, at $Z = 250$ cm and a 20 dB SNR (black curve), nearly 50\% of the estimates exhibit errors exceeding 20 cm. This trend confirms a direct relationship between increased vertical distance and reduced positioning accuracy, especially under low SNR conditions, due to the accumulated propagation loss and geometric dilution. Hence, the proposed 3D positioning system is more sensitive to vertical variations at low SNRs, emphasizing the importance of signal strength in high-precision localization.
\begin{figure}[b]
    \centering   \includegraphics[width=1\linewidth]{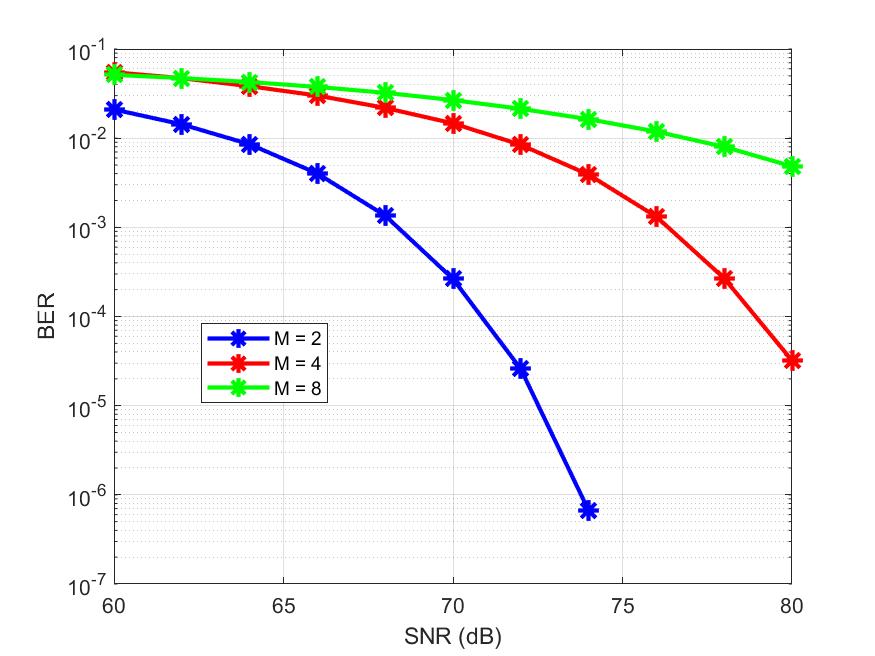}
    \caption{Effect of PD location relative to the room center on BER performance.}
    \label{fig:13}
\end{figure}

The effect of user location relative to the center on BER performance is also investigated, following the earlier observation of location dependence in positioning accuracy. Fig.~\ref{fig:13} illustrates the BER performance at three different receiver locations on the same plane, corresponding to different distances from the room center. At each SNR level, \(10^6\) bits were transmitted to the receivers positioned at these locations. The results show that the receiver closest to the center achieves significantly better BER performance across all SNR levels than the receivers positioned farther away. This shows that user proximity to the room center enhances communication reliability and positioning accuracy due to improved channel conditions and reduced path loss.
\begin{figure}[t]
    \centering    \includegraphics[width=1\linewidth]{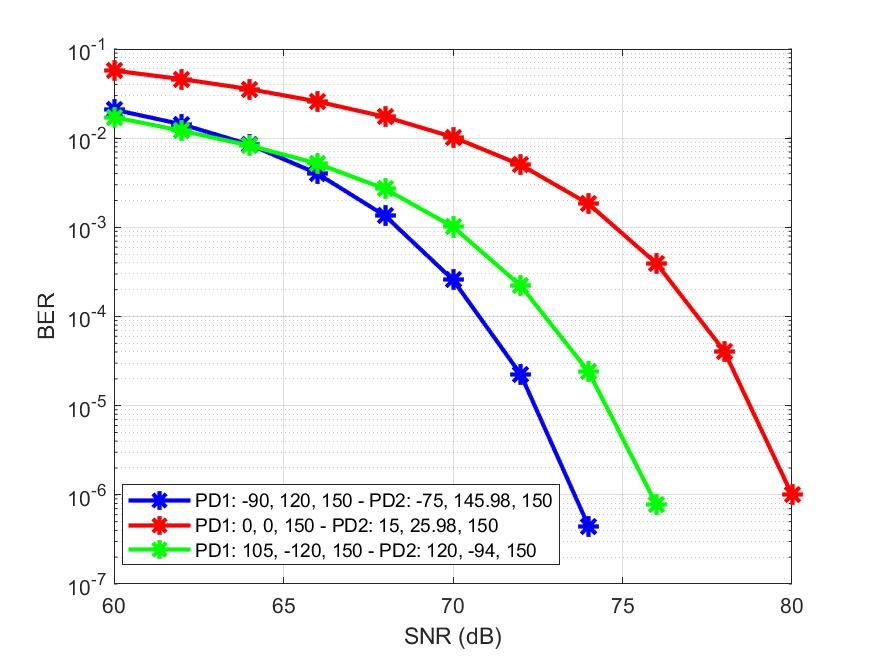}
    \caption{BER performances of M-ary PAM based SM.}
    \label{fig:12}
\end{figure}

Following the positioning performance evaluation, the communication performance of the proposed system is assessed. Fig.~\ref{fig:12} illustrates the BER performance of the M-ary PAM-based SM system with \(M = 2\), \(M = 4\), and \(M = 8\) at a selected receiver position of \((-2.5, 1.5, 0)\). In this simulation, \(10^6\) information bits were transmitted from randomly selected LEDs to the receiver across SNR levels ranging from 60~dB to 80~dB. The results indicate that the BER decreases steadily as the SNR increases for all modulation orders. It can be observed that the system with \(M=2\) achieves the best performance, with the BER dropping below \(10^{-6}\) at high SNRs. As the modulation order increases to \(M=4\) and \(M=8\), the BER performance degrades due to the reduced Euclidean distance between constellation points, as expected. These results confirm that the proposed system can achieve highly reliable communication under high SNR conditions while maintaining the ability to perform joint communication and positioning simultaneously.

\section{Conclusion}
A joint communication and positioning algorithm based on VLC has been proposed to address the need for high-precision indoor localization and reliable connectivity. The unified framework leverages SM and M-ary PAM, enabling simultaneous data transmission and positioning with minimal system complexity. Simulation results confirm that the method achieves sub-centimeter accuracy in both 2D and 3D positioning scenarios at high SNR levels, with 3D errors increasing with receiver height under low SNR due to geometric dilution and signal attenuation. Communication performance also remains robust, achieving BERs below $10^{-6}$ for $M=2$ PAM above 70~dB SNR. Furthermore, both positioning and communication accuracy are maximized when the receiver is near the geometric center of the LED layout. These findings highlight the potential of VLC systems as a practical enabler for integrated localization and communication in future 6G indoor networks.




\bio{}
\noindent \textbf{Tarik Anil Leblebici} (Student Member, IEEE) received the B.Sc. degree in Electrical and Electronics Engineering with high honors from Kadir Has University, Istanbul, Turkey, in 2024. His research interests include visible light communication, joint communication and positioning, and next-generation wireless systems.
\endbio

\medskip

\noindent\textbf{Sumeyra Hassan} (Graduate Student Member, IEEE) received her B.Sc. degree, ranking third in her university, in 2019 and her M.Sc. degree in 2021 from Kadir Has University. She is currently pursuing a Ph.D. degree while working as a Research Assistant in the Electrical and Electronics Engineering Department at Kadir Has University. Her research interests include next-generation wireless communications, optical communications, ISAC, and physical layer security.

\medskip

\noindent \textbf{Erdal Panayirci} (Life Fellow IEEE) received a Diploma Engineering degree in electrical engineering from Istanbul Technical University, Istanbul, Türkiye, and a Ph.D. degree in electrical engineering and system science from Michigan State University, East Lansing, MI, USA. He is currently a Professor at the Electrical and Electronics Engineering Department at Kadir Has University, Istanbul. He holds a Visiting Research Collaborator position with the Department of Electrical Engineering, Princeton University, Princeton, NJ, USA. He has published extensively in leading scientific journals and international conferences and coauthored the book \textit{Principles of Integrated Maritime Surveillance Systems} (Kluwer Academic, 2000). His research interests include communication theory, synchronization, advanced signal processing techniques, and their applications to wireless electrical, underwater, and optical communications. Prof. Panayirci was an Editor of the \textit{IEEE Transactions on Communications} in Synchronization and Equalization from 1995 to 2000. He is a member of the National Academy of Sciences of Turkey. He served as a member of the IEEE Fellow Committee from 2005 to 2008 and from 2018 to 2020 and as a member of the IEEE GLOBECOM/ICC Management and Strategy Standing Committee from 2017 to 2021. He is currently a member of the IEEE ComSoc Awards Standing Committee. He served as the Technical Program Co-Chair, the General Co-Chair, and the Technical Program Chair for several IEEE ICC, IEEE PIMRC, and IEEE WCNC conferences.

\medskip

\noindent\textbf{H. Vincent Poor} (Life Fellow, IEEE) received the Ph.D. degree in EECS from Princeton University in 1977. From 1977 until 1990, he was on the faculty of the University of Illinois at Urbana-Champaign. Since 1990, he has been on the faculty at Princeton, where he is currently the Michael Henry Strater University Professor. During 2006 to 2016, he served as the dean of Princeton’s School of Engineering and Applied Science, and he has also held visiting appointments at several other universities, including most recently at Berkeley and Caltech. His research interests are in the areas of information theory, machine learning, and network science, and their applications in wireless networks, energy systems, and related fields. Among his publications in these areas is the book \textit{Machine Learning and Wireless Communications} (Cambridge University Press, 2022). Dr. Poor is a member of the National Academy of Engineering and the National Academy of Sciences and is a foreign member of the Royal Society and other national and international academies. He received the IEEE Alexander Graham Bell Medal in 2017.

\end{document}